\documentclass[%
aip,
 amsmath,amssymb,
 preprint,%
]{revtex4-1}

\usepackage{graphicx}% Include figure files
\usepackage{dcolumn}% Align table columns on decimal point
\usepackage{xcolor}
\usepackage{bm}% bold math
\usepackage[utf8]{inputenc}
\usepackage[T1]{fontenc}
\usepackage{mathptmx}
\usepackage{etoolbox}

\makeatletter
\def\@email#1#2{%
 \endgroup
 \patchcmd{\titleblock@produce}
  {\frontmatter@RRAPformat}
  {\frontmatter@RRAPformat{\produce@RRAP{*#1\href{mailto:#2}{#2}}}\frontmatter@RRAPformat}
  {}{}
}%
\makeatother
\begin{document}

\preprint{AIP/123-QED}

\title[Injection dynamics in spin-wave active ring oscillator (SWARO)]{Injection dynamics in spin-wave active ring oscillator (SWARO)}
% Force line breaks with \\
\author{Anirban Mukhopadhyay}
\email{ee19d022@smail.iitm.ac.in}
\affiliation{Department of Electrical Engineering,
Indian Institute of Technology,
Chennai 600036, India.}%

\author{Ihor I. Syvorotka}%
\affiliation{Department of Crystal Physics and Technology, Scientific Research Company “Electron-Carat”, Lviv 79031, Ukraine}%
\affiliation{Department of Semiconductor Electronics, Lviv Polytechnic National University, Lviv 79013, Ukraine.}%
% \altaffiliation[also at]{Department of Semiconductor Electronics, Lviv Polytechnic National University, Lviv 79013, Ukraine.}

\author{Anil Prabhakar}
\affiliation{Department of Electrical Engineering,
Indian Institute of Technology,
Chennai 600036, India.}
%\homepage{http://www.Second.institution.edu/~Charlie.Author.}

\date{\today}% It is always \today, today,

\begin{abstract}
We investigated injection locking in spin-wave active ring oscillators (SWAROs) operating in the multi-mode regime. By applying external RF signals with varying frequency and power, we identified the locking behavior of individual modes and extracted the total locking ranges from spectral measurements. The results show asymmetric evolution of the lower and upper locking boundaries with drive power for the lower-frequency SWARO modes, while the highest-frequency mode exhibits nearly symmetric behavior. A maximum locking range of over 11~MHz is observed at a drive power of -10~dBm. \textcolor{black}{To interpret these results, we develop an Adler-like model that captures the dependence of the locking range on drive power, showing good agreement for the higher-frequency modes. For the lowest-frequency mode, however, the model underestimates the locking range at low drive and saturates at high drive power levels, while the experimental range increases monotonically, indicating the influence of multi-mode interactions. These findings establish SWARO as a useful platform for exploring injection phenomena in spin-wave ring systems with delayed feedback and motivate the development of extended injection models that account for multi-mode dynamics.} 
\end{abstract}

\maketitle

%%%%%%%%%%%%%%%
% Introduction
%%%%%%%%%%%%%%%
\section{Introduction}
Spin-wave active ring oscillators (SWAROs) are a class of magnetic microwave oscillators that comprises a magnetic film acting as a spin-wave delay line and a variable gain unit.
The eigenmodes are resonantly excited when the round-trip delay is an integer multiple of 2$\pi$ and losses are fully compensated \cite{WU2010163}.
The resonance condition can also be influenced by several parameters, including the delay line thickness, saturation magnetization, applied magnetic field, and the specific spin-wave configuration.
The SWARO acts as a multi-mode resonator when the losses are not fully compensated.

% The free-running SWARO presents itself as an attractive test bed for studying various nonlinear phenomena, such as soliton generation \cite{Kalinikos1998, Kalinikos1999, Kalinikos2000}, chaos \cite{Wu2009, Hagerstrom2009, Kondrashov2010, Hagerstrom2011}
% , Hagerstrom2012, Kondrashov2015, Ustinov2021,}, 
% , foldover of nonlinear eigen-modes \cite{Janantha2017}, bistability \cite{Vitko2021, Vitko2022}, etc.
Researchers reported various nonlinear dynamics in the driven SWARO system, such as excitation of dark spin-wave solitons by injecting the ring resonator with two periodic signals whose frequencies coincide with ring eigenmodes \cite{Ustinov2024biharmonic}.
\textcolor{black}{Alternatively, an external RF signal was shown to suppress the chaotic spectral output of the SWARO \cite{Grishin2008620}.} 
In addition to this, chaotic microwave pulses were also generated by driving the SWARO with pulse-modulated RF signal \cite{Grishin2013321, Grishin2013gen}.

In magnetic microwave oscillators, the injection locking of spin-torque and spin-Hall nano-oscillators has been the subject of extensive investigation, with particular emphasis on phase synchronization and linewidth reduction \cite{Rippard2005, Quinsat2011, Tortarolo2018}.
Furthermore, injection dynamics were studied under a variety of external excitations, including surface acoustic waves \cite{Sravani_2022}, radio-frequency magnetic fields \cite{Singh2018}, and injected alternating currents \cite{Cao2014}.
\textcolor{black}{Recently, a two-stage cavity magnonic injection-locked amplifier achieved a lock-in range exceeding 10~MHz for input powers between –30 and –12~dBm \cite{kim_two_stage_2025}.}
Injection locking in SWAROs introduces additional complexity due to multi-mode competition and delayed feedback. 
Our work, therefore, complements earlier studies by providing a systematic investigation of injection locking in a multi-mode spin-wave delay-line oscillator.

\textcolor{black}{In this work, a static magnetic field was applied in the plane of a YIG film acting as the delay-line and oriented perpendicular to the spin-wave propagation direction in the magnetostatic surface spin-wave (MSSW) configuration. The delay-line was closed via a feedback loop with variable gain.} A continuous RF signal of variable power was injected into the oscillator, and the injection-locking range of each excited mode was experimentally determined as a function of the drive power. The measured ranges were subsequently compared against an Adler-like analytical model.

%%%%%%%%%%%%%%%%%%%%%%%%%%%%%%%%%%%%%%%%%%%%%%%%%%
% Transmission spectroscopy of MSSW delay line
%%%%%%%%%%%%%%%%%%%%%%%%%%%%%%%%%%%%%%%%%%%%%%%%%%
\section{Transmission spectroscopy of MSSW delay line}
\begin{figure}[!htbp]
    \centering
    \includegraphics[width=0.65\linewidth]{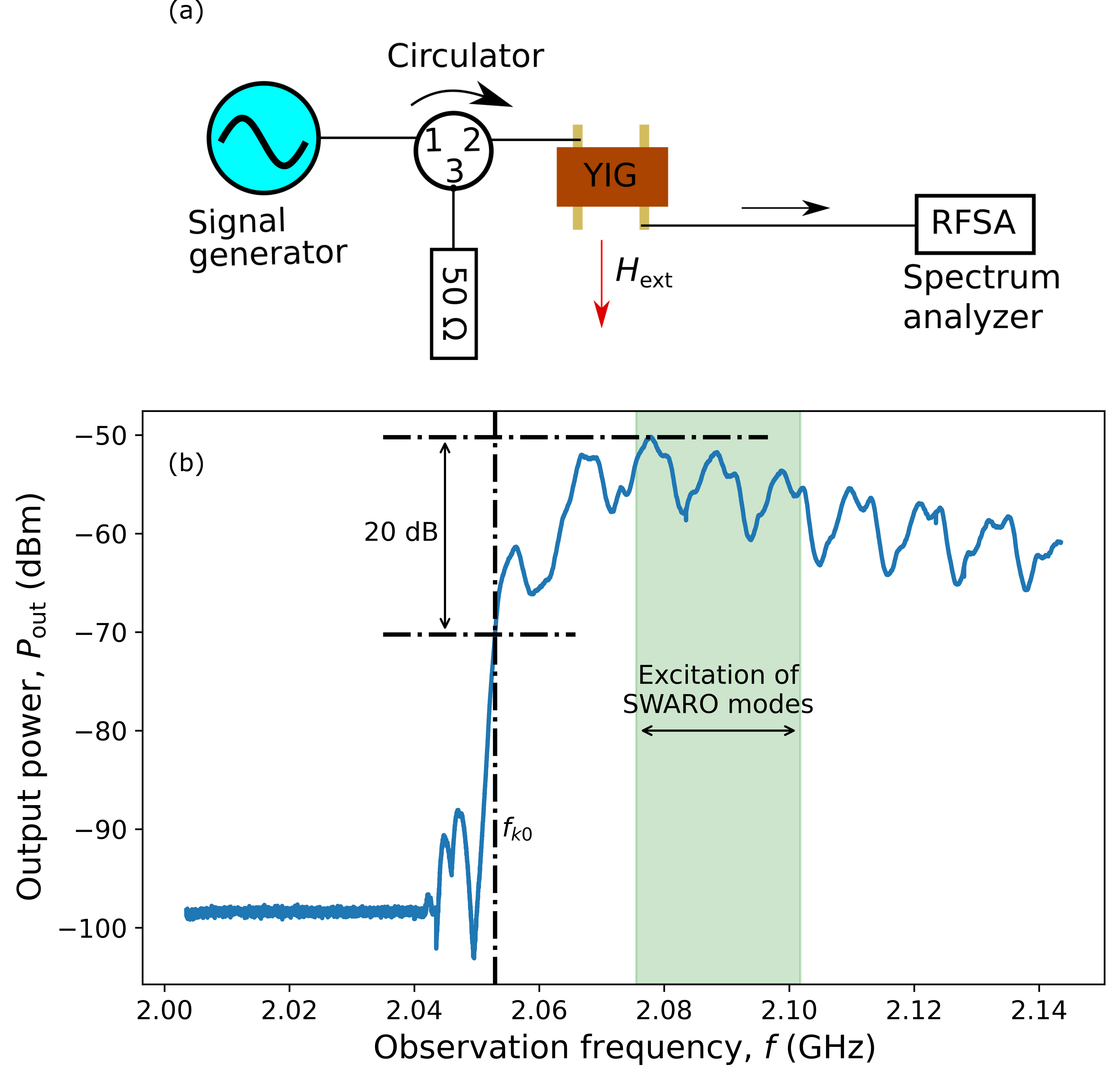}
    \caption{(a) Transmission spectroscopy setup for detecting MSSWs propagating through the YIG film. (b) Transmission characteristics of MSSWs through the YIG delay line. The input power, $P_\text{in}$ = -40~dBm. The lower cut-off frequency of the MSSW manifold, $f_{k\text{0}}$ = 2.05291~GHz, which corresponds to an effective field $H_\text{eff}$ = 21.27~kA/m (estimated via Eq.~(\ref{eq:lower_limit_mssw_manifold})).}
    \label{fig:swtx_setup+mssw_tx_spectrum}
\end{figure}

Figure~\ref{fig:swtx_setup+mssw_tx_spectrum}(a) shows the experimental setup used for measuring transmission characteristics of the YIG delay line.
The YIG film was kept on a pair of $\mu$-strip line antennas, separated by a distance of $l_\text{sw} \approx \text{9~mm}$.
The film has a size 18~mm $\times$ 14~mm $\times$ 6.9~$\mu$m, with a saturation magnetization, $M_\text{s}$ = 138.46~kA/m.
An effective magnetic field of $H_\text{eff} = \text{21.27~kA/m}$ saturated the film along the width of the YIG film.
A signal of strength of -40~dBm was fed to the excitation antenna, and the signal frequency was swept from 2 to 2.13~GHz in steps of 10~kHz.
The receiving antenna converted the MSSWs into RF signals and sent them to a spectrum analyzer.
The transmission characteristic of the YIG delay line is plotted in Fig.~\ref{fig:swtx_setup+mssw_tx_spectrum}(b). The MSSW dispersion relation can be represented as
\begin{align}
\label{eq:mssw_disp}
f^2 = f_\text{M}^2\left[\frac{H_\text{eff}}{M_\text{s}}\left(\frac{H_\text{eff}}{M_\text{s}} + 1\right) + \frac{(1-\exp(-2kt))}{4}\right],
\end{align}
where, $t$ is the YIG film thickness, $f_\text{M}=|\gamma|\mu_\text{0}M_\text{s}/\left(2\pi\right)=\text{4.87379~GHz}$.
$\gamma$ and $\mu_\text{0}$ are the gyromagnetic ratios for the electron and free-space magnetic permeability, respectively. 
In the limit $k \rightarrow 0$, Eq.~(\ref{eq:mssw_disp}) yields $H_\text{eff}$ as
\begin{align}
\label{eq:lower_limit_mssw_manifold}
    H_\text{eff} &= M_\text{s} \left(-\frac{1}{2} + \sqrt{\frac{1}{4} + \left(\frac{f_{k\text{0}}}{f_\text{M}}\right)^2}\right).
\end{align}
The lower limit of the MSSW manifold, $f_{k\text{0}}$ in Eq.~(\ref{eq:lower_limit_mssw_manifold}) was obtained from Fig.~\ref{fig:swtx_setup+mssw_tx_spectrum}(b) by assuming that at this frequency, the transmitted power decreased by 20~dB from its maximum.
This YIG delay line, along with a fixed-gain amplifier and a variable step attenuator, constructed a SWARO, and we analyzed the driven dynamics of SWARO in the frequency domain.

%%%%%%%%%%%%%%%%%%%%%%%%%%%%%%%%%%%%%%%%%%%%%%%%%%
% Driven dynamics of SWARO
%%%%%%%%%%%%%%%%%%%%%%%%%%%%%%%%%%%%%%%%%%%%%%%%%%
\section{Driven dynamics of SWARO}
\begin{figure}[!htbp]
    \centering
    \includegraphics[width=0.9\linewidth]{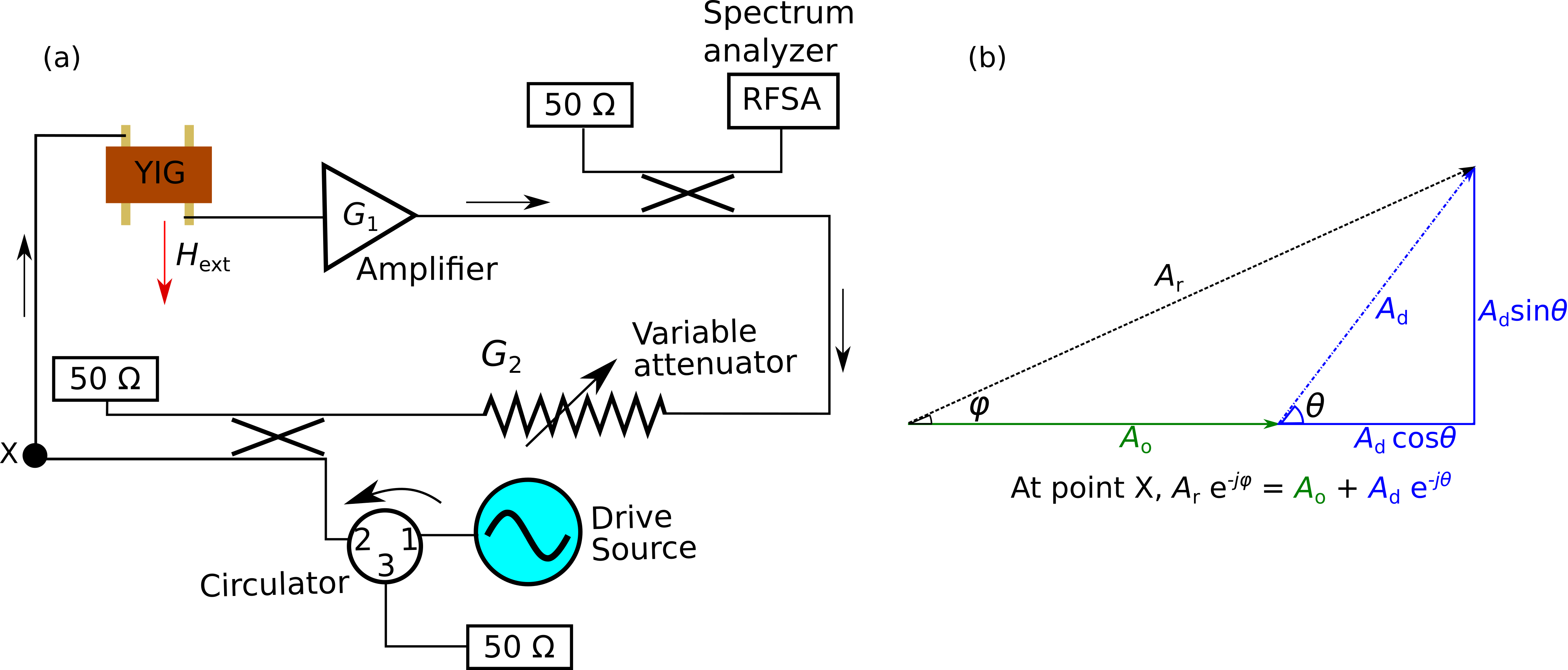}
    \caption{(a) Circuit diagram of spin-wave active ring oscillator (SWARO) with RF generator. (b) Phasor diagram for signals circulating in the driven SWARO circuit.}
    \label{fig:swaro_ckt_w_rf_gen}
\end{figure}
\begin{figure}[!htbp]
    \centering
    \includegraphics[width=0.75\linewidth]{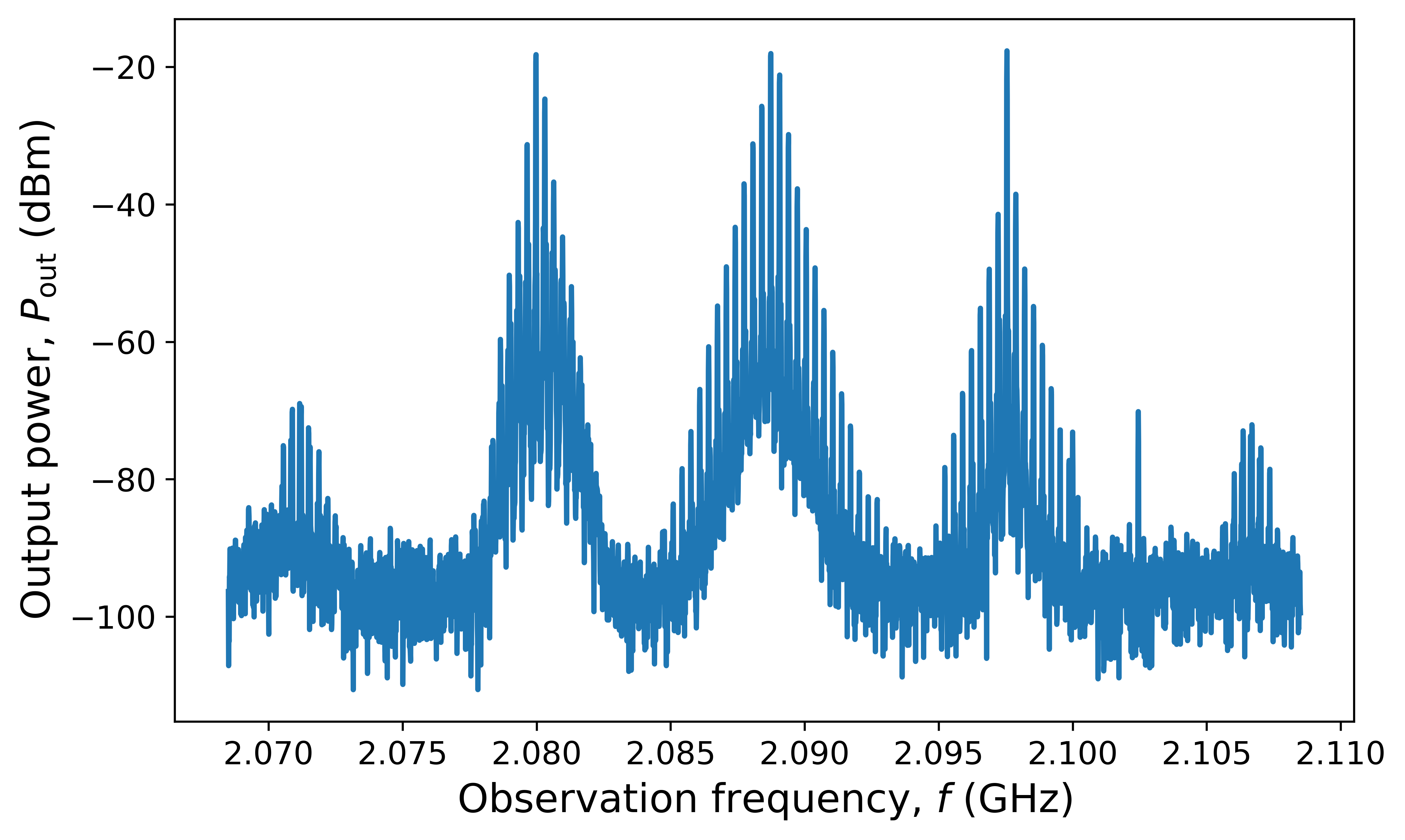}
    \caption{Spectral output of free-running SWARO at $G_\text{2}$ = -8~dB. The SWARO modes are $f_\text{1}$ = 2.07998~GHz,  $f_\text{2}$ = 2.08874~GHz,  $f_\text{3}$ = 2.09755~GHz. The sidebands in the neighborhood of the SWARO mode are separated by a distance of 330~kHz.}
    \label{fig:free_run_swaro_spec}
\end{figure}
Figure~\ref{fig:swaro_ckt_w_rf_gen} shows the circuit of a SWARO with a drive oscillator. 
The SWARO consisted of the YIG film, a constant-gain amplifier, a variable step attenuator, and two bi-directional couplers.
The couplers were used to inject continuous GHz signals and extract approximately 1\% of the power generated in the ring.
The RF spectrum analyzer recorded the frequency response of the SWARO with and without the external drive.
The analyzer's observation frequency window spanned between 2.0685~GHz and 2.1084~GHz. The RBW of the analyzer was set to 10~kHz.
We set the $G_2$ = -8~dB, which excited three SWARO modes at $f_\text{1}$ = 2.07998~GHz,  $f_\text{2}$ = 2.08874~GHz,  and $f_\text{3}$ = 2.09755~GHz, shown in Fig.~\ref{fig:free_run_swaro_spec}.
The separation between the modes are $\Delta f_{12} = \text{8.76~MHz}$ and $\Delta f_{23} = \text{8.81~MHz}$.
Grishin~\emph{et al.} reported the emergence of frequency combs in the spectral output of a spin-wave active ring oscillator, attributing their origin to the three-wave decay process. In their study, the comb spacing was found to lie between 270 and 300 kHz \cite{Grishin2013gen}. 
\textcolor{black}{We also observe frequency combs, with a spacing of approximately 330 kHz in Fig.~\ref{fig:free_run_swaro_spec} and attribute them to a similar three-wave mixing mechanism.}

\begin{figure}
    \centering
    \includegraphics[width=\linewidth]{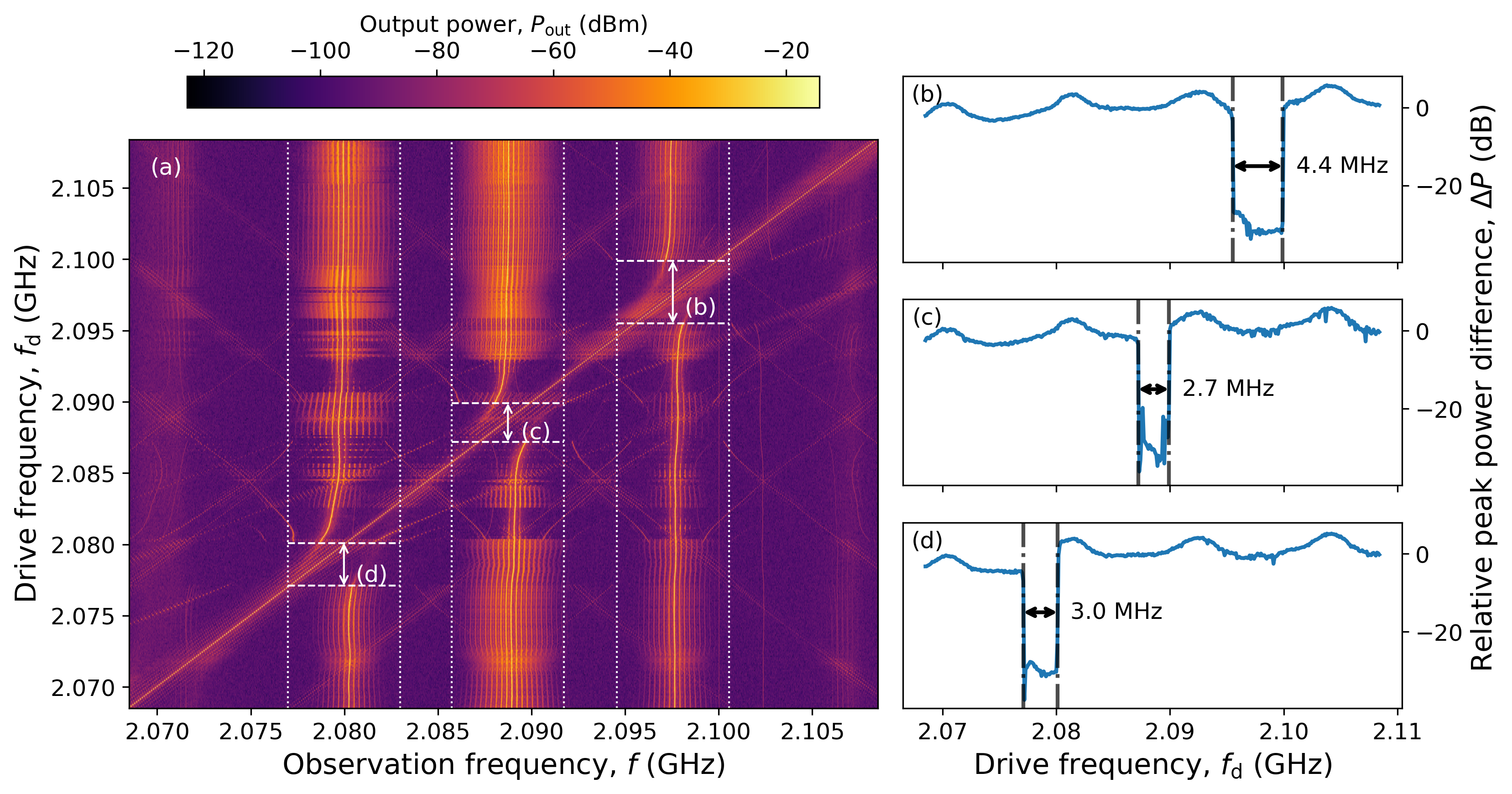}
    \caption{(a) Spectral output of SWARO at $P_\text{d} = \text{-15~dBm}$, plotted as a function of observation frequency $f$ and drive frequency $f_\text{d}$. Vertical dotted lines mark the $\pm$~3~MHz windows around each free-running mode. Dashed horizontal lines indicate the extracted injection-locking ranges for modes 1–3. (b–d) Relative peak power difference $\Delta P$ as a function of $f_\text{d}$ for each mode, showing sharp dips that define the upper and lower locking boundaries. The corresponding bounds are highlighted by horizontal dashed lines in (a). Suppression of sidebands in neighboring modes is evident when a given mode is pulled toward the drive frequency.}
    \label{fig:spec_driven_swaro_-15dbm}
\end{figure}

Next, we injected the SWARO with the drive signal of varying power and frequency.
The drive power, $P_\text{d}$, was varied from -30~dBm to -10~dBm with a step-size of 1~dB.
We swept the drive frequency, $f_\text{d}$, from 2.0685~GHz to 2.1084~GHz with a step size of 100~kHz.
The spectral output of the ring oscillator was captured for each combination of power and frequency.
Fig.~\ref{fig:spec_driven_swaro_-15dbm}(a) plots surface map $P_\text{out}$ in a parameter space constituted by $(f, f_\text{d})$, at $P_\text{d} = -15~\mathrm{dBm}$.
Each row in the surface map corresponds to a spectral output from the SWARO at a fixed $f_\text{d}$.
In optical oscillators, injection locking is commonly identified by a side-mode suppression ratio exceeding 20~dB \cite{Hui1991, Hiraoka2021, Ratkoceri2021}. In spintronic nano-oscillators, the injection-locked state is typically indicated by a reduction in linewidth \cite{Quinsat2011, Rajabali2023}. In this work, we estimated the injection locking range from the spectral maps using the relative peak power difference $\Delta P$ as the criterion. The estimation method was as follows:
\begin{enumerate}
\item We extracted the power values along the diagonal line, i.e., the output power at the drive frequency $P_\text{out}(f_\text{d}, f_\text{d})$.
\item Windows of width $w = \text{6~MHz}$ centered at each SWARO mode were defined, i.e. $|f- f_i| \le w,\:\:\forall i = 1, 2, 3$. The dotted vertical line pairs indicate three windows in Fig.~\ref{fig:spec_driven_swaro_-15dbm}.
\item In each frequency window, we selected the power value corresponding to the strongest non-driven spectral component, represented as $P^\prime$.
\item We estimated the relative peak power difference $\Delta P = P^\prime - P_\text{out}(f_\text{d}, f_\text{d})$.
\end{enumerate}
In Figs.~\ref{fig:spec_driven_swaro_-15dbm}(b–d), $\Delta P$ is plotted as a function of $f_\text{d}$. 
Each curve exhibits a pronounced dip with steeply defined edges, and these two edges correspond to the lower and upper bounds, $f_\text{L,U}$ of the injection locking region. These bounds are also indicated by dashed horizontal lines in Fig.~\ref{fig:spec_driven_swaro_-15dbm}(a). 
The total locking range is defined as $\Delta f = f_\text{U} - f_\text{L}$.
Furthermore, when the first mode (second mode) was pulled towards the higher drive frequency, the sidebands associated with the second and third (first and third) modes were suppressed.
\begin{figure}
    \centering
    \includegraphics[width=0.75\linewidth]{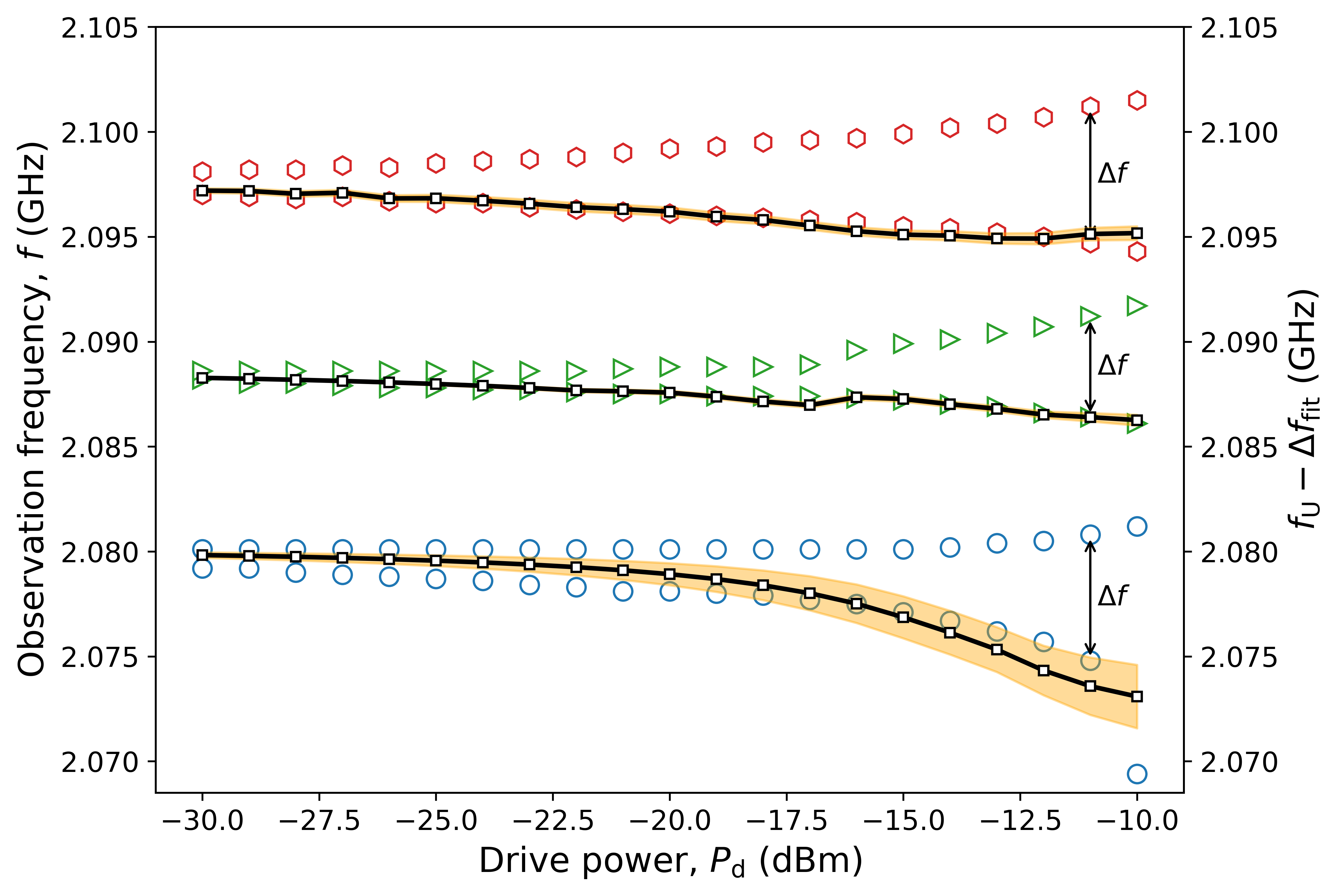}
    \caption{The experimentally estimated upper and lower limits of injection lock-in are shown for three SWARO modes $f_\text{1}$, $f_\text{2}$, and $f_\text{3}$. $f_\text{U} - \Delta f_\text{fit}$ as a function of drive power $P_\text{d}$ is plotted. $\Delta f_\text{fit}$ is prediction from the theoretical model given in Eq.~\ref{eq:adler_model}. The orange regions denote the 95\% confidence intervals of the fit.}
    \label{fig:up_low_lim_lockin+fit}
\end{figure}
Fig.~\ref{fig:up_low_lim_lockin+fit} presents the variation of $f_\text{L,U}$ with increasing $P_\text{d}$ for the three excited SWARO modes. 
It can be observed that increasing the drive power affects the upper and lower frequency limits differently; in particular, for modes 1 and 2, $f_\text{L}$ shifts at a noticeably faster rate than $f_\text{U}$. 
By contrast, for mode 3, $f_\text{L}$ and $f_\text{U}$ exhibit nearly symmetric variation with $P_\text{d}$. Notably, at $P_\text{d} = \text{-10~dBm}$, the total locking range $\Delta f$ exceeds 11~MHz for mode 1.

%%%%%%%%%%%%%%%%%%%%%%%%%%%%%%%%%%%%%%%%%%%%%%%%%%
% Theoretical model
%%%%%%%%%%%%%%%%%%%%%%%%%%%%%%%%%%%%%%%%%%%%%%%%%%
\section{Theoretical model}
\textcolor{black}{Adler originally developed an injection-locking model for LC oscillators under sinusoidal injection currents \cite{Adler1946, Hong2019}. In the present work, an Adler-like model with low computational overhead is developed to describe the injection-locking phenomena observed in the SWARO.} 
We assume that the oscillator operates in a single-mode regime and that the drive frequency is tuned in close proximity to the oscillator mode.
The signals traveling through the circuit are represented in their phasor forms and shown as a vector diagram in Fig.~\ref{fig:swaro_ckt_w_rf_gen}(b) \cite{Mirzaei2006}.
A drive source injects a GHz signal of frequency $f_\text{d}$, amplitude $A_\text{d}$, and power $P_\text{d} = |A_\text{d}|^2$ into the SWARO, and the SWARO is locked to the drive source. 
$A_\text{o}$ denotes the amplitude of the free-running SWARO output, while $A_\text{r}$ corresponds to the resultant signal.
$\varphi$ and $\theta$ are the phase angles between $A_\text{o}$ and $A_\text{d}$ and $A_\text{o}$ and $A_\text{r}$, respectively.
The round-trip phase accumulated by MSSWs propagating through the YIG delay line can be expressed as \cite{prabhakar2009spin}
\begin{align}
    \label{eq1:mssw_disp}
    \Phi(f) \approx k(f)l_\text{sw} &= \frac{-l_\text{sw}}{2t}\ln\left[1 - \frac{4\left(f^2 - f_\text{H}^2 - f_\text{H}f_\text{M}\right)}{f_\text{M}^2}\right],
    \nonumber\\
    &= \frac{-l_\text{sw}}{2t}\ln\left[1 - \chi(f)\right],\nonumber\\
    &\approx \frac{l_\text{sw}}{2t}\left(\chi(f) + \frac{\chi(f)^2}{2}\right),
\end{align}
for $|\chi| \ll 1$, where $\chi(f) = \frac{4\left(f^2 - f_\text{H}^2 - f_\text{H}f_\text{M}\right)}{f_\text{M}^2}$, $f_\text{H} = \gamma \mu_0 H_\text{eff} / ( 2 \pi)$, $f_\text{M} = \gamma \mu_0 M_\text{s} / ( 2 \pi)$, $l_\text{sw}$ is the distance between the antenna pair and $k$ is the MSSW wave number.
\textcolor{black}{The approximation in Eq.~(\ref{eq1:mssw_disp}) yields a mean absolute error of 0.0005~rad/mm.}
We assumed that $G = G_1 + G_2$ had a value that allowed the SWARO to operate as a single-mode oscillator. 
Figure~\ref{fig:swaro_ckt_w_rf_gen}(b) describes the relationship between $\theta$ and $\varphi$ which can be expressed as
\begin{align}
    \label{eq2:phi_phasor_diag}
    \tan(\varphi) &= \frac{\beta \sin\theta}{1 + \beta \cos\theta},\nonumber\\
    \therefore\:\: \varphi &= \tan^\text{-1}\left(\frac{\beta \sin\theta}{1 + \beta \cos\theta}\right),
\end{align}
where the injection strength, $\beta = \frac{A_\text{d}}{A_\text{o}} = \sqrt{\frac{P_\text{d}}{P_\text{o}}}$.
$P_\text{d}$ and $P_\text{o}$ are in linear scale.

Since $f_\text{d}$ is assumed to be close to $f_i$, the round-trip phase accumulation at $f_\text{d}$, $\Phi(f_\text{d})$, can be approximated by a first-order Taylor expansion about $f = f_i$.
\begin{align}
    \label{eq3:phi_taylor_expansion}
    \Phi(f_\text{d}) &\approx \Phi(f_i) + (f_\text{d} - f_i) \frac{d\Phi}{df}(f_i)\nonumber,\\
    \varphi &= \Phi(f_\text{d}) - \Phi(f_i) \approx  l_\text{sw} (f_\text{d} - f_i) \frac{dk}{df}(f_i),
\end{align}
where, $\Phi(f_i)$ is an integer multiple of 2$\pi$ and $f_i$ = SWARO mode frequency.
\textcolor{black}{In the next step, we equate Eq.~(\ref{eq2:phi_phasor_diag}) and (\ref{eq3:phi_taylor_expansion}) to obtain an expression of the total locking range,}
\begin{align}
\label{eq:adler_model}
    \Delta f_\text{fit} = |f_\text{L}^\text{fit} - f_i| + |f_\text{U}^\text{fit} - f_i| &= \frac{2}{l_\text{sw} \frac{dk}{df}(f_i)} \tan^{-1}\left(\frac{\beta \sin \theta}{\beta \cos \theta + 1}\right),
\end{align}
where, $f_\text{L,U}^\text{fit}$ = lower and upper limits of injection locking predicted by the model, $P_\text{o},\:\:\theta$ are fitting parameters.
Figure~\ref{fig:up_low_lim_lockin+fit} fits the experimental data to the analytical model described in Eq.~(\ref{eq:adler_model}). 
\textcolor{black}{We note that the assumptions made in the derivation differ from the actual experimental conditions.} In practice, the SWARO operates in a multi-mode regime, and the drive frequency is not always confined to the vicinity of a single oscillator mode. These factors introduce additional nonlinear interactions and mode competition, which are not captured by the simplified Adler-like model.

\textcolor{black}{To evaluate the model, confidence bands were calculated. The narrow confidence bands for the higher-frequency modes indicate good agreement with the data. However, for the lowest-frequency SWARO mode, the model underestimates the locking range at low $P_\text{d}$ and fails to capture the trend at high $P_\text{d}$, where the arc-tangent dependence, inherent to the Adler-like model, predicts saturation while the experimental locking range increases monotonically. These results highlight the limitations of the existing model.}

%%%%%%%%%%%%%%%%%%%%%%%%%%%%%%%%%%%%%%%%%%%%%%%%%%
% Conclusion
%%%%%%%%%%%%%%%%%%%%%%%%%%%%%%%%%%%%%%%%%%%%%%%%%%
\section{Conclusion}
In conclusion, this work investigates the injection-locking phenomena in SWAROs operating in the multi-mode regime. Injection signals with varying power and frequency were applied to the oscillator, and the corresponding output spectra were recorded. The total injection-locking range was determined by identifying the sharp edges of the dips in the $\Delta P$ versus $f_\text{d}$ curves. The results reveal that the lower bound of the locking range shifts at a faster rate with increasing $P_\text{d}$ than the upper bound for the lower-frequency modes, while for the third mode, the variation is nearly symmetric. Notably, the total locking range exceeds 11~MHz for the lowest-frequency mode at $P_\text{d} = \text{-10~dBm}$. \textcolor{black}{An Adler-like model was developed to describe the dependence of the locking range on drive power. While it captures the behavior of the higher-frequency SWARO modes, deviations in the lowest mode underscore its limited applicability. Although perturbation projection (PPV) and time-synchronous impulse sensitivity function (ISF) frameworks could potentially overcome these shortcomings, they involve considerable mathematical complexity \cite{Bhansali2009, Hong2019}. Future work will extend these advanced frameworks to more accurately capture multi-mode injection-locking dynamics.}
 
\section*{Author Declarations}
\subsection*{Acknowledgments}
We thank N. Bilanuik and D. D. Stancil for the micro-strips, the Department of Science and Technology for support vide sanction order IWT/UKR/P-28/2018, and acknowledge the bilateral Indian–Ukrainian S\&T cooperation project (M/11-2020). \textcolor{black}{A.M. is also thankful to the Mphasis F1 Foundation for financial support.}
\subsection*{Conflict of Interest}
The authors have no conflicts to disclose.
\subsection*{Author contributions}
\textbf{Anirban Mukhopadhyay}: Conceptualization(equal); Data curation (lead); Formal analysis (equal); Investigation (equal); Methodology (equal); Software(lead); Validation (equal); Visualization(equal); Writing – Original Draft (equal); Writing - Review \& Editing (equal). \textbf{Ihor I. Syvorotka}: Project Administration (supporting); Resources (supporting). \textbf{Anil Prabhakar}: Conceptualization(equal); Data curation (supporting); Formal analysis (equal); Investigation (equal); Methodology (equal); Project Administration (lead); Resources (lead); Supervision (lead); Validation (equal); Visualization(equal); Writing – Original Draft (equal); Writing - Review \& Editing (equal).

\section*{Data Availability Statement}
The data that support the findings of this study are available from the corresponding author upon reasonable request.

\bibliography{REF}% Produces the bibliography via BibTeX.

\end{document}